\pdfoutput=1
\documentclass[10pt,journal,compsoc]{IEEEtran}

\usepackage{booktabs}
\usepackage{graphicx}
\usepackage{tabularx}
\usepackage{xcolor}
\usepackage{xspace}
\usepackage{url}
\usepackage{tikz}
\usetikzlibrary{arrows.meta,positioning,calc,fit,shapes.geometric}
\usepackage{listings}
\usepackage{colortbl}
\definecolor{diffdel}{HTML}{CF222E}
\definecolor{diffadd}{HTML}{1A7F37}
\definecolor{diffdelbg}{HTML}{FFEBE9}
\definecolor{diffaddbg}{HTML}{E6FFEC}
\definecolor{diffhead}{HTML}{EAEEF2}
\lstdefinestyle{codeline}{basicstyle=\ttfamily, columns=fullflexible, keepspaces=true}
\newcommand{\codenote}[1]{\leavevmode{\rmfamily\itshape\color{black!60}#1}}
\newcommand{\diffdel}{\rowcolor{diffdelbg}\textcolor{diffdel}{\texttt{-}}}
\newcommand{\diffadd}{\rowcolor{diffaddbg}\textcolor{diffadd}{\texttt{+}}}
\newcommand{\diffctx}{\texttt{\phantom{+}}}
\newsavebox{\diffbox}
\newcommand{\diffhead}[2]{\multicolumn{2}{>{\columncolor{diffhead}}p{\dimexpr\columnwidth-6.8pt\relax}}{\sffamily #1\hfill\texttt{#2}}}

\ifCLASSOPTIONcompsoc
  \usepackage[nocompress]{cite}
\else
  \usepackage{cite}
\fi
\usepackage[hidelinks]{hyperref}

\newcommand{\finding}[2]{\vspace{2pt}\noindent\fbox{\parbox{0.97\columnwidth}{\textbf{Answer to #1:} #2}}\vspace{3pt}}
\definecolor{idlink}{HTML}{1F4FA0}
\newcommand{\idref}[2]{\href{#1}{\textcolor{idlink}{#2}}}
\newcommand{\pgcommit}[1]{\idref{https://github.com/postgres/postgres/commit/#1}{\texttt{#1}}}
\newcommand{\duckcommit}[1]{\idref{https://github.com/duckdb/duckdb/commit/#1}{\texttt{#1}}}
\newcommand{\mariacommit}[1]{\idref{https://github.com/MariaDB/server/commit/#1}{\texttt{#1}}}
\newcommand{\crdbcommit}[1]{\idref{https://github.com/cockroachdb/cockroach/commit/#1}{\texttt{#1}}}
\newcommand{\sqlitecommit}[1]{\idref{https://sqlite.org/src/info/#1}{\texttt{#1}}}

\newcommand{\VfiveApolloSqliteTimed}{167{,}183}
\newcommand{\VfiveApolloSqliteAbove}{564}
\newcommand{\VfiveApolloSqliteBelowMinimum}{561}
\newcommand{\VfiveApolloPgReports}{2{,}020}
\newcommand{\VfiveMicroShareLow}{30}
\newcommand{\VfiveMicroShareHigh}{44}
\newcommand{\VfiveSpeedtestMax}{4.5}
\newcommand{\VfivePgbenchMax}{0.4}
\newcommand{\VfiveMicroAlertEffect}{23}
\newcommand{\VfiveGeneratorApplicable}{81}
\newcommand{\VfiveBaselineApplicable}{18}
\newcommand{\VfiveGeneratorFamily}{15}
\newcommand{\VfiveGeneratorFamilyCases}{7}
\newcommand{\VfiveAmoebaSets}{3{,}126}
\newcommand{\VfiveAmoebaReports}{27}

\begin{document}

\title{From Code Changes to Slowdowns:\\
An Empirical Study of Performance Regressions in Open-Source Database Kernels}

\author{Xueli~Sun and Shuangxiang~Kan%
\IEEEcompsocitemizethanks{%
  \IEEEcompsocthanksitem X.~Sun is with Nanjing University of Posts and
  Telecommunications, Nanjing, China.\protect\\
  E-mail: sunxueli@njupt.edu.cn
  \IEEEcompsocthanksitem S.~Kan is with Singapore Management University,
  Singapore.\protect\\
  E-mail: sxkan@smu.edu.sg (Corresponding author)
  \IEEEcompsocthanksitem This work is supported by the National Natural Science Foundation of China (Nos.~62402241, 62372248, and 62272244), the Natural Science Foundation of the Jiangsu Higher Education Institutions of China (No.~24KJD520006), Natural Science Research Start-up Foundation of Recruiting Talents of Nanjing University of Posts and Telecommunications (No.~NY223189), Natural Science Foundation of Jiangsu Province (No.~BK20200753), Jiangsu Postdoctoral Science Foundation Funded Project (No.~2021K096A), and the Open Project of State Key Laboratory of Internet Architecture (No.~HLW2025MS24).}}

\IEEEtitleabstractindextext{%
\begin{abstract}
Database kernels change continuously, and a change that leaves every query
result correct can still make a workload far slower. Because results stay
correct, result-checking tests cannot reveal such regressions, and although
several detectors target database performance problems, little is known about
the regressions that code changes introduce. We present, to our
knowledge, the first cross-system study of code-change-induced performance
regressions in database kernels. Scanning over 2.1 million commits and related
reports of nine open-source systems, from embedded SQLite to
distributed TiDB, we identify 668 publicly confirmed regression families in
kernel code and analyze where they occur, how changes cause them, and how
completely they can be traced to code. We find that regressions span the kernel:
query optimization is the largest module (27.4\%), but execution and storage
together account for more (36.8\%), and the leading module differs by
architecture. Extra work on a hot path, plan changes, and concurrency or
coordination changes cause 61.8\% of families. Over 95\% of families map to
changed code, but only 43.7\% identify the introducing change, a subset that
over-represents plan changes. Replaying 22 regressions on historical builds, we
evaluate nine configurations of four detectors, including APOLLO and AMOEBA, and
three developer benchmarks. Most timing alerts are slowdowns that the
regression's fix does not remove; the detectors reach only seven regressions and
the benchmarks none, mainly because their workloads never contain the trigger.
These results call for change-directed testing that varies the workload
dimensions a change puts at risk.
\end{abstract}


}

\maketitle
\IEEEdisplaynontitleabstractindextext
\IEEEpeerreviewmaketitle

\ifCLASSOPTIONcompsoc
\IEEEraisesectionheading{\section{Introduction}\label{sec:introduction}}
\else
\section{Introduction}
\label{sec:introduction}
\fi

\IEEEPARstart{D}{atabase} management systems (DBMSs) are a critical component of
modern data-intensive applications~\cite{jung2019apollo}. More than one
trillion SQLite databases are estimated to be in active
use~\cite{sqlite2026mostdeployed}, and PostgreSQL and MySQL were the databases
most used by professional developers in the 2025 Stack Overflow
survey~\cite{stackoverflow2025survey}. For the
applications built on a DBMS, performance matters alongside correctness.
Response time determines how quickly an application can take in new
information and act on it~\cite{jung2019apollo}, so applications set latency
targets, size their hardware, and tune their queries on the assumption that a
query that is fast today will stay fast after the next upgrade. In
dependability terms, a system's function covers performance as well as
functionality, and correct results delivered too late constitute a timing
failure~\cite{avizienis2004taxonomy}. Database kernels, however, change
continuously, and a change to the optimizer, the executor, the storage engine,
or synchronization can slow down a workload without altering a single query
result. We call such a slowdown a \emph{code-change-induced performance
regression}. Because results stay correct, result-checking tests
cannot reveal it.

Public reports show how severe such regressions can be. A planner heuristic for
star-schema queries made a user's SQLite query that took 1.3\,s on release
3.46.1 run for more than an hour on
3.47.2~\cite{sqliteforum2024regression,sqliteforum2025views}; a costlier hash
function and InnoDB functions that were no longer inlined made join queries
about 18\% slower on MySQL 8.0.33 than on 8.0.28~\cite{mysqlbug111538}; and
PostgreSQL semaphores that shared cache lines caused contention on many-CPU
systems until release 10.5 padded them~\cite{pg2018release105}. Regressions are
also costly to handle after release. Users who encounter one after an upgrade
often cannot report it, because the triggering query is complex and depends on
their own database, and developers must then spend considerable effort
localizing the cause~\cite{jung2019apollo}. Plan regressions are painful enough
in production that commercial systems store earlier plans and revert to
them~\cite{ding2018planstitch}, and Meta's pre-production testing platform must
detect regressions as small as 0.01\%~\cite{chow2024servicelab}.

Catching a regression before release is difficult, however. Its trigger can
involve SQL shape, schema, statistics, data scale, cache state, concurrency,
topology, or configuration; in the optimizer alone, cardinality estimates are
routinely wrong and can lead to poor join orders~\cite{leis2015optimizers}. Even
with the right workload, the slowdown must be separated from measurement noise,
which in cloud environments can exceed 100\%~\cite{laaber2019cloud}. Database
projects combine unit tests, project-specific performance tests, and workloads
such as TPC-C or Sysbench, but replaying broad benchmarks at commit frequency is
expensive, and a fixed workload exercises only a small part of a kernel's
behavior. Research detectors such as APOLLO~\cite{jung2019apollo},
AMOEBA~\cite{liu2022amoeba}, CERT~\cite{ba2024cert}, and more recent
tools~\cite{wu2025puppy,wu2025hulk,ba2026queryzen} find performance bugs
automatically, but most derive their oracles from optimizer behavior and
examine one version at a time.

What is missing is empirical knowledge of the regressions themselves. To our
knowledge, no study has characterized the regressions that code changes introduce across database architectures, so two questions that shape
detector design remain open. The
first is what these regressions look like: a population dominated by optimizer
changes would favor plan-sensitive SQL generation, whereas one rich in
hot-path, storage, and synchronization regressions would require scale, cache,
I/O, and concurrency dimensions and latency or throughput oracles. The second is
what ground truth exists: evaluating a commit-level detector requires the
introducing change, but many reports identify only the fix, and inferring the
former from the latter, as SZZ-style algorithms do, does not always recover
it~\cite{sliwerski2005szz,rodriguezperez2020bugs,lyu2024szz}.

This paper presents an empirical study of publicly confirmed code-change-induced
performance regressions in nine open-source database kernels: SQLite,
PostgreSQL, DuckDB, TiDB, MySQL, MariaDB, CockroachDB, OceanBase, and
ClickHouse. We scan their public histories, which comprise more than 2.1 million
commits and check-ins, together with their issue trackers, mailing lists, and
release notes. Our unit of analysis is a regression \emph{family}: all public
artifacts that share one code cause, so that a report, its backport, and its
fix count once. We separate regressions acknowledged or fixed upstream,
regressions reproduced locally, and performance costs that maintainers
explicitly accepted for another goal. We ask four research
questions:

\begin{description}[\IEEEsetlabelwidth{RQ4}]
  \item[RQ1] \emph{Where?} Which kernel modules contain publicly confirmed
  code-change-induced performance regressions, and how does the composition
  vary across database architectures?
  \item[RQ2] \emph{Why?} Which code-change mechanisms cause the observed
  regressions?
  \item[RQ3] \emph{What evidence?} How completely can a regression family be
  linked to an introducing/fixing change, public evidence, and source
  locations?
  \item[RQ4] \emph{Detected?} Can existing detectors and developer benchmarks
  expose these regressions, and why do they miss?
\end{description}

\noindent RQ1--RQ3 are answered from the corpus. RQ4 is answered by a dynamic
evaluation on 22 evaluation cases derived from the corpus and replayed on exact
historical builds. Our main
findings are as follows.

\begin{itemize}
  \item \textbf{Regressions are not mainly an optimizer problem (RQ1).} Of
  668 kernel families, query optimization is the largest module
  (27.4\%), but query execution and storage I/O together account for more
  (36.8\%), and the leading module differs by architecture. Plan-only testing
  cannot cover this population.
  \item \textbf{Three mechanisms dominate (RQ2).} Extra work on a hot path,
  plan selection or transformation changes, and concurrency or coordination
  changes cause 61.8\% of families. Plan changes mostly stay in the optimizer, but hot-path work spans every module, so no single performance test covers them.
  \item \textbf{Introducing changes are often unknown (RQ3).} Over 95\% of
  families can be localized to changed files, functions, or hunks, but only
  43.7\% identify the introducing change, and this subset over-represents plan
  changes and under-represents hot-path work. Commit-level evaluations must
  choose their ground truth explicitly and must not relabel a fix as the cause.
  \item \textbf{Existing detectors rarely reach the studied regressions
  (RQ4).} Nine detector configurations, including APOLLO and
  AMOEBA, and three developer benchmarks raise many timing alerts, but
  re-measuring each alert across the fixing commit attributes alerts to the
  studied family in only \VfiveGeneratorFamilyCases{} of the 22 cases, and the
  benchmarks attribute none. Most misses occur because workloads never contain the triggering construct;
timer resolution explains few of them.
  \item \textbf{Tests should vary the suspected trigger (Section~\ref{sec:implications}).} The causal
  evidence maps a changed code region to the workload dimensions and oracles
  needed to expose it. Data scale and SQL shape are the most common dimensions,
  but neither dominates, so generation should be conditioned on the change
  rather than replay one workload after every commit.
\end{itemize}

This paper makes three contributions:

\begin{itemize}
  \item A corpus of 668 regression families from nine database kernels,
  collected by a source-native protocol adapted to each community's
  workflow and linked, where available, to introducing and fixing changes,
  source locations, triggers, and test hypotheses.
  \item A cross-system characterization of affected modules, change mechanisms,
  confirmation status, and evidence completeness, together with the requirements these data impose on change-risk analysis, directed test generation,
  performance oracles, and leakage-free evaluation.
  \item A dynamic evaluation of nine detector configurations and three
  developer benchmarks on 22 evaluation cases replayed on exact historical
  builds of SQLite, DuckDB, PostgreSQL, and TiDB. It attributes alerts to
  regression families by re-measuring them across the fixing commit and
  separates misses from inapplicable or incompatible cases.
\end{itemize}

Sections~\ref{sec:related} and~\ref{sec:design} cover related work and the
study design, Section~\ref{sec:results}
answers RQ1--RQ3, and Section~\ref{sec:dynamic} reports the dynamic evaluation
(RQ4). Sections~\ref{sec:discussion} and~\ref{sec:threats} discuss implications
and threats to validity, and Section~\ref{sec:conclusion} concludes.

\section{Background and Related Work}
\label{sec:related}

\subsection{Performance Bug and Regression Studies}

Performance regressions are slowdowns or resource-efficiency losses caused by a
software change relative to a relevant baseline. General software studies show
that performance bugs frequently arise from inefficient computation, repeated
work, synchronization, and misunderstanding of library or system
behavior~\cite{jin2012performance}. Repository studies compare performance bugs
with other bugs and show that they differ in how they are discovered, reported,
and fixed~\cite{zaman2012qualitative,nistor2013discovering}. Han and Yu examine
113 performance bugs from Apache, MySQL, and Firefox with a focus on the role of
configuration~\cite{han2016configurable}. In long-running systems,
aging-related bugs cause performance to decline over execution time, and recent
work predicts them across projects~\cite{xu2025aging}. These studies
characterize performance bugs regardless of whether a prior change introduced
them. Database kernels add
optimizer decisions, physical operators, storage policies, and distributed
coordination to this space.

Closer to our unit of analysis, Chen and Shang benchmark 1,126 Hadoop and 135
RxJava commits to identify and analyze performance-regression-introducing
changes~\cite{chen2017exploratory}. Ren et al.\ trace the performance evolution
of Linux core operations across 41 kernel versions and attribute most observed
slowdowns to a small number of kernel changes~\cite{ren2019linux}. For
distributed storage and computing systems, TaxPerf classifies 99 performance
bugs from five systems into six root-cause categories~\cite{li2023taxperf}.
These studies either measure regressions in a few non-database code bases or
analyze performance bugs without identifying the introducing change. To our
knowledge, no prior study characterizes publicly confirmed code-change-induced
regressions across database kernels while separating introducing from fixing
revisions.

\subsection{Performance Regressions in Practice}

Plan regressions are a recognized production problem: commercial systems store
earlier plans and revert to them~\cite{ding2018planstitch}, and regressions on
some queries still hinder the adoption of learned
optimizers~\cite{weng2024eraser}. Storage engines raise their own diagnosis
problems, as performance issues in MongoDB's WiredTiger
illustrate~\cite{fedorova2018wiredtiger}. Outside databases, large operators
invest heavily in catching small regressions. Meta's pre-production testing
platform must detect regressions as small as 0.01\%~\cite{chow2024servicelab},
and its production monitoring tracks about 800,000 time series to catch
regressions caused by code or configuration changes~\cite{yoon2024fbdetect}.
One year of Firefox performance testing raised 17,989 alerts that had to be
triaged~\cite{besbes2025mozilla}. Performance testing in open-source projects,
by contrast, lacks accepted best practices~\cite{leitner2017perftesting}.
Measurement noise is a further obstacle: in cloud environments, benchmark
variability ranges from negligible to more than 100\%, and slowdowns of 10\% or
less become detectable when the compared versions run on the same machines in
randomized order~\cite{laaber2019cloud}. Our dynamic evaluation follows this
design by interleaving versions on one host (Section~\ref{sec:dynamic-design}).

\subsection{Performance Testing of Database Systems}

APOLLO combines workload generation, version comparison, and statistical
debugging; it reported ten regressions in SQLite and PostgreSQL, seven
acknowledged by developers~\cite{jung2019apollo}. AMOEBA mutates a query into an
equivalent form and treats a substantial timing divergence as a potential
performance bug~\cite{liu2022amoeba}. CERT generates queries whose estimated
cardinality should increase under a controlled transformation and flags plan
costs that violate the expected relation; its evaluation found 13 issues across
three DBMSs, nine confirmed~\cite{ba2024cert}. PUPPY compares a fully optimized
plan with plans in which selected optimizations are disabled; it reported 62
performance degradation bugs in five DBMSs, 54 confirmed as previously
unknown~\cite{wu2025puppy}. Hulk searches for data-volume-sensitive performance
cliffs caused by cost-based optimization and reported 135 anomalies in six
DBMSs, 129 confirmed as new bugs~\cite{wu2025hulk}. QueryZen flips
optimization-related code branches and treats a significantly faster flipped
execution as a performance issue; it found 21 previously unknown issues in four
DBMSs~\cite{ba2026queryzen}. Recent work also studies query performance
regressions caused by practical index-tuning decisions~\cite{wu2026index}.

Most of these techniques derive their oracles from optimizer behavior, such as
equivalent rewrites, cardinality estimates, disabled optimizations, or forced
branches, and all except APOLLO examine one version at a time instead of a code
change. Our study complements them by characterizing what a change-directed detector should cover, including execution, storage, and
coordination regressions that optimizer-derived oracles are not designed to
expose.

\subsection{Change-Aware Performance Regression Testing}

Commit-level approaches aim to avoid running every performance test after every
change. PerfScope analyzes the performance risk of a code change to prioritize
performance regression testing~\cite{huang2014perfscope}. PerfRanker
prioritizes performance regression tests for collection-intensive
software~\cite{mostafa2017perfranker}, and PerfJIT predicts at the test level
whether a commit introduces a performance regression~\cite{chen2022perfjit}. In
database continuous integration, MongoDB applies change-point detection to
continuous performance measurements~\cite{daly2020changepoint}, and BIPeC
combines Bayesian inference with PELT change-point detection to identify
regressions in SAP HANA test measurements~\cite{lyu2024bipec}. These approaches
select or prioritize existing tests, or detect regressions after measurement;
they do not establish which workload dimensions a test must vary to trigger a
given class of database regression. Our corpus provides such evidence (Section~\ref{sec:implications}) and
task-specific ground truth for evaluating commit-level detection.

Such ground truth needs the introducing change, whereas many public reports
identify only the fixing commit. Inferring the former from the latter, as the
SZZ algorithm does~\cite{sliwerski2005szz}, assumes that the lines changed by a
fix introduced the bug. This assumption does not always
hold~\cite{rodriguezperez2020bugs}, and an evaluation on developer-labeled pairs
in the Linux kernel shows that some introducing commits cannot be traced this
way at all~\cite{lyu2024szz}. We therefore never infer an introducing revision
from a fix (Section~\ref{sec:design}).

\subsection{Database Testing and Bug Studies}

Database testing has advanced rapidly for correctness. Pivoted Query Synthesis
constructs queries around a pivot row~\cite{rigger2020pqs}; NoREC compares an
optimized query with a non-optimizing reference form~\cite{rigger2020norec}; and
Query Plan Guidance steers generation toward new plans~\cite{ba2023qpg}.
Coverage-guided fuzzing is also widely used to find DBMS bugs, and recent work
makes its coverage feedback more accurate and adds test-case
trimming~\cite{li2025dbmsfuzzing}. These
methods motivate plan- and transformation-aware test generation, but their
oracles do not identify performance degradation.

A recent study of 777 bugs in three relational DBMSs provides a broad taxonomy
of database failures~\cite{liu2025rdbmsbugs}. BugForge collects 37,632 reports
from four systems to support DBMS testing~\cite{li2026bugforge}. Outside
databases, a study of 3,555 bug reports from three deep learning frameworks
classifies faults by their triggering conditions~\cite{du2023faulttriggers}. Unlike these general bug corpora, our study focuses on performance degradation
introduced by code changes and records, for each regression, the causal chain,
the performance mechanism, and the workload dimensions needed to trigger it. We also cover nine
systems and explicitly separate accepted trade-offs from repaired regressions.
Studies of performance anti-patterns in ORM-based web
applications~\cite{yang2018orm} address a different layer: the application code
that issues queries rather than the kernel that executes them.

\section{Study Design}
\label{sec:design}

\subsection{System Selection and Snapshot}

Table~\ref{tab:systems} summarizes the corpus. We selected systems to cover
embedded (SQLite), traditional client/server (PostgreSQL, MySQL, MariaDB),
analytical/vectorized (DuckDB, ClickHouse), and distributed SQL or
NewSQL-style systems (TiDB, CockroachDB, OceanBase). They also differ in
implementation language and public workflow, making a single issue-only mining
strategy inappropriate. The snapshots were frozen between August 28 and 30,
2026.

\begin{table}[t]
\caption{Unified study corpus. ``All'' includes confirmed families, adjacent
components, and terminal evidence-insufficient records.}
\label{tab:systems}
\centering
\small
\begin{tabular}{lrrrr}
\toprule
System & All & Confirmed & Kernel & Adjacent \\
\midrule
SQLite       & 190 & 91  & 87  & 4  \\
PostgreSQL   & 149 & 149 & 143 & 6  \\
DuckDB       & 35  & 35  & 35  & 0  \\
TiDB         & 165 & 165 & 136 & 29 \\
MySQL        & 103 & 103 & 102 & 1  \\
MariaDB      & 41  & 41  & 41  & 0  \\
CockroachDB  & 33  & 33  & 32  & 1  \\
OceanBase    & 29  & 29  & 27  & 2  \\
ClickHouse   & 66  & 66  & 65  & 1  \\
\midrule
Total        & 811 & 712 & 668 & 44 \\
\bottomrule
\end{tabular}
\end{table}

\subsection{Source-Native Collection}

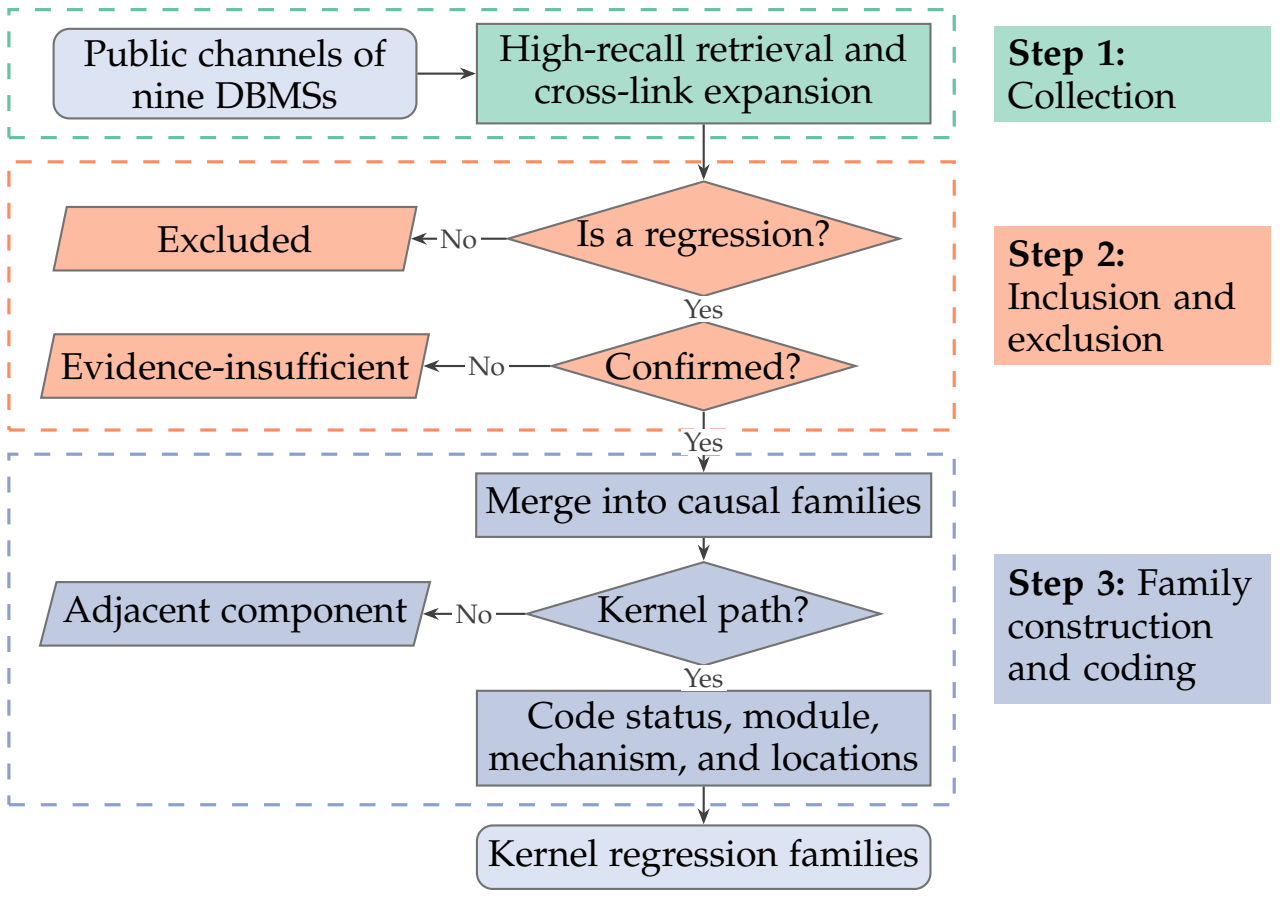
\begin{figure}[t]
\centering
\definecolor{fA}{HTML}{66C2A5}
\definecolor{fB}{HTML}{FC8D62}
\definecolor{fC}{HTML}{8DA0CB}
\definecolor{fT}{HTML}{DCE3F2}
\begin{tikzpicture}[
  font=\scriptsize, >={Stealth[length=3pt]},
  fterm/.style={draw=black!55, fill=fT, rounded corners=3pt, align=center,
    minimum width=3.0cm, minimum height=4.2mm, inner sep=1.8pt},
  fproc/.style={draw=black!55, fill=#1, align=center, minimum width=3.0cm,
    minimum height=4.2mm, inner sep=1.8pt},
  fdec/.style={draw=black!55, fill=#1, diamond, aspect=3.4, align=center,
    inner sep=0.3pt},
  flab/.style={draw=black!55, fill=#1, trapezium, trapezium left angle=72,
    trapezium right angle=108, trapezium stretches body, align=center,
    minimum width=2.4cm, minimum height=4.2mm, inner sep=1.2pt},
  fgrp/.style={draw=#1, dashed, semithick, inner sep=0pt},
  fstp/.style={fill=#1, text width=1.65cm, align=left, inner sep=2.5pt,
    execute at begin node={\hyphenpenalty10000\relax}},
  far/.style={->, black!75},
  fyn/.style={font=\tiny, fill=white, inner sep=0.6pt},
]
\node[fproc=fA!55] (ret) {High-recall retrieval and\\cross-link expansion};
\coordinate (lx) at ($(ret.center)+(-3.1cm,0)$);
\node[fterm, minimum width=2.4cm] (start) at (lx |- ret) {Public channels of\\nine DBMSs};
\coordinate (s1) at (ret |- start.south);
\node[fdec=fB!60, below=4mm of s1] (d1) {Is a regression?};
\node[fdec=fB!60, below=1.5mm of d1] (d2) {Confirmed?};
\node[fproc=fC!55, below=4mm of d2] (merge) {Merge into causal families};
\node[fdec=fC!55, below=1.5mm of merge] (d3) {Kernel path?};
\node[fproc=fC!55, below=1.5mm of d3] (code) {Code status, module,\\mechanism, and locations};
\node[fterm, below=2.5mm of code] (end) {Kernel regression families};
\node[flab=fB!60] (l1) at (lx |- d1) {Excluded};
\node[flab=fB!60] (l2) at (lx |- d2) {Evidence-insufficient};
\node[flab=fC!55] (l3) at (lx |- d3) {Adjacent component};
\node[fgrp=fA, fit={($(l2.west |- start.north)+(-2.5mm,1.2mm)$) ($(ret.east |- start.south)+(1.5mm,-1.2mm)$)}] (g1) {};
\node[fgrp=fB, fit={($(l2.west |- d1.north)+(-2.5mm,1.2mm)$) ($(ret.east |- d2.south)+(1.5mm,-1.2mm)$)}] (g2) {};
\node[fgrp=fC, fit={($(l2.west |- merge.north)+(-2.5mm,1.2mm)$) ($(ret.east |- code.south)+(1.5mm,-1.2mm)$)}] (g3) {};
\node[fstp=fA!55, right=2.5mm of g1.east] {\textbf{Step 1:} Collection};
\node[fstp=fB!60, right=2.5mm of g2.east] {\textbf{Step 2:} Inclusion and exclusion};
\node[fstp=fC!55, right=2.5mm of g3.east] {\textbf{Step 3:} Family construction and coding};
\draw[far] (start) -- (ret);
\draw[far] (ret) -- (d1);
\draw[far] (d1) -- node[fyn] {Yes} (d2);
\draw[far] (d2) -- node[fyn] {Yes} (merge);
\draw[far] (merge) -- (d3);
\draw[far] (d3) -- node[fyn] {Yes} (code);
\draw[far] (code) -- (end);
\draw[far] (d1) -- node[fyn] {No} (l1);
\draw[far] (d2) -- node[fyn] {No} (l2);
\draw[far] (d3) -- node[fyn] {No} (l3);
\end{tikzpicture}
\caption{Corpus construction. Evidence-insufficient records and
adjacent-component families remain in our ledger.}
\label{fig:pipeline}
\end{figure}

Figure~\ref{fig:pipeline} shows the collection pipeline. For each project, we
first enumerated its native public sources: Git or Fossil history, issue and pull
request trackers, comments, mailing lists or forums, release notes, and, where
relevant, CommitFest or bug-tracker records. High-recall queries combined
performance terms (e.g., \emph{regression}, \emph{slower}, \emph{throughput}),
causal terms (e.g., \emph{introduced}, \emph{since}, \emph{bisect}), and
resolution terms (e.g., \emph{fix}, \emph{revert}). We followed cross-source
links and commit identifiers rather than treating a keyword hit as evidence.

Across the nine histories (Table~\ref{tab:coverage}), we scanned 2,108,623 Git
commits or Fossil check-ins.
Project-specific collectors yielded 123,994 candidate rows for adjudication.
We do not add these quantities to issue, comment, or release counts, because
their units overlap and differ by source. They describe the search boundary, not
regression prevalence.

\begin{table}[t]
\caption{History coverage and adjudicated candidate rows.}
\label{tab:coverage}
\centering
\small
\begin{tabular}{lrr}
\toprule
System & Commits/check-ins & Candidates \\
\midrule
SQLite      & 36{,}848  & 1{,}603 \\
PostgreSQL  & 104{,}376 & 43{,}597 \\
DuckDB      & 95{,}325  & 2{,}925 \\
TiDB        & 326{,}263 & 8{,}127 \\
MySQL       & 420{,}215 & 9{,}945 \\
MariaDB     & 317{,}683 & 9{,}490 \\
CockroachDB & 227{,}897 & 24{,}316 \\
OceanBase   & 44{,}835  & 864 \\
ClickHouse  & 535{,}181 & 23{,}127 \\
\midrule
Total       & 2{,}108{,}623 & 123{,}994 \\
\bottomrule
\end{tabular}
\end{table}

\subsection{Inclusion, Exclusion, and Status}

We define a candidate as a \emph{code-change-induced performance regression}
when a source change causes a measurable or explicitly reported degradation in
latency, throughput, resource use, or scalability relative to an earlier
revision. A family is confirmed when at least one of the following holds:

\begin{enumerate}
  \item an upstream developer statement, fix, or revert explicitly connects a
  code change to the degradation;
  \item the upstream evidence identifies affected revisions and our differential
  experiment reproduces the degradation; or
  \item maintainers explicitly acknowledge a performance cost accepted for
  correctness, security, memory use, or another engineering objective.
\end{enumerate}

We exclude ordinary optimizations without a prior regression, benchmark noise,
pure correctness failures, environment-only failures, and slowdown reports that
cannot be connected to a code change. The 99 unresolved SQLite records
are terminal evidence-insufficient exclusions.

The 668-family main corpus contains 541 upstream-confirmed regressions, seven
locally reproduced regressions, and 120 acknowledged trade-offs. We retain the
last group because it describes performance consequences that commit-level
testing should surface, but report it separately: an accepted cost is not a
``fixed performance bug.'' Forty-four confirmed families in connectors, clients,
or other adjacent components remain in our ledger but are excluded from
kernel distributions.

\subsection{Family Construction and Coding}

Our unit of analysis is a causal family, not a public artifact. Reports,
backports, release-note entries, reverts, and fixes are merged when their links,
commit identifiers, and descriptions establish a shared code cause. Each family
retains all source references, permitting the merge to be audited.

We map project-specific labels into two normalized dimensions. The \emph{primary
module} describes where the regression manifests (e.g., query optimization,
query execution, storage I/O, or concurrency/transaction). The \emph{change
mechanism} describes how behavior worsened (e.g., extra hot-path work, plan
selection, or coordination). Original project labels and multi-module context
are preserved even though aggregate tables use one primary value. The codebook
was refined iteratively, then applied with deterministic normalization rules so
the aggregate dataset can be regenerated. Every family also receives a targeted
test hypothesis derived from its causal evidence.

For code mapping, introducing and fixing revisions are resolved against the
read-only source mirrors. We record changed files, enclosing functions when
recoverable, and diff-hunk context. A fixing revision is never inferred to be the
introducing revision. This distinction determines which subsets are valid for
change-prediction experiments.

\subsection{Integrity Checks and Analysis}

The unified build performs machine-checkable invariants before emitting results:
global record identifiers must be unique; per-system source and output counts
must agree; all confirmed kernel records must have a title, module, and mechanism;
and the partitions must be complete. The current audit verifies
$811=668+44+99$: main kernel families, adjacent-component families, and terminal
evidence-insufficient records, respectively. It also executes each of the nine
project-level audits. Aggregate values in this paper are regenerated from the
machine-readable ledger rather than transcribed from case narratives.

Our analysis is descriptive. The collected families are an observed public
population under heterogeneous reporting processes, not a random sample from an
unknown defect population. We therefore report counts and within-system
proportions and avoid null-hypothesis tests whose sampling assumptions would
mislead. Cross-system differences identify coverage requirements rather than
rank systems.

\subsection{Experimental Environment}
\label{sec:environment}

All experiments in this paper run on one Ubuntu 22.04 server with two 96-core
AMD EPYC 9654 processors and 1\,TB of memory. Each revision is built
from upstream source as an optimized build with GCC 11.4, Clang 14, or
Go 1.19, and all revisions of a case use the same toolchain and flags. Each
revision runs in a fresh, private database instance, and timed processes
are pinned to fixed physical cores.

\section{Study Results}
\label{sec:results}

\subsection{RQ1: Affected Kernel Modules}

\begin{figure}[t]
\centering
\includegraphics[width=\columnwidth]{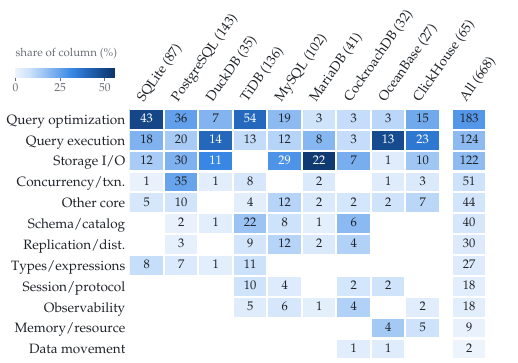}
\caption{Primary modules of the 668 kernel families. Numbers are family
counts; shading is the share within each column, a system or the whole
corpus (All). Blank cells are zero.}
\label{fig:modules}
\end{figure}

Figure~\ref{fig:modules} gives the primary-module distribution, overall and by
system. Query optimization is the largest individual category (183, 27.4\%), so
optimizer behavior is a major source of regressions, but it does not account for
a majority. Query execution (124) and storage I/O (122) together contribute 246
families (36.8\%), and adding concurrency and transactions raises this share to
44.5\%. A detector that observes only plan changes would miss much of the
corpus.

The composition also reflects architecture. Optimizer regressions account for
43/87 SQLite families (49.4\%) and 54/136 TiDB families (39.7\%). PostgreSQL has
a broader concentration in concurrency/transactions (35/143, 24.5\%) and
storage (30/143, 21.0\%). Storage is the largest category in MySQL (29/102,
28.4\%) and MariaDB (22/41, 53.7\%). Query execution leads in DuckDB (14/35,
40.0\%), OceanBase (13/27, 48.1\%), and ClickHouse (23/65, 35.4\%).

Grouping the systems by architecture (Section~\ref{sec:design}) shows that the
leading module changes with architecture. In the embedded engine, the optimizer
accounts for about half of the families (43/87, 49.4\%). The three client/server systems regress most often in storage
I/O (81/286, 28.3\%), followed by query optimization (20.3\%), query execution
(14.0\%), and concurrency and transactions (12.9\%). In the two analytical
engines, query execution leads (37/100, 37.0\%). In the three distributed systems, most regressions lie in the SQL layer, led by
query optimization (60/195, 30.8\%), query execution (14.9\%), and
schema/catalog (14.4\%), whereas the replication/distributed module contributes
only 13 families (6.7\%), fewer than the 17 replication families of the
client/server systems. A detector that targets several systems must adapt its
testing dimensions to the code region and the architecture, and a detector for
distributed systems needs SQL-level workloads as much as cluster-level ones.

Three cases show why. In MariaDB, change \idref{https://jira.mariadb.org/browse/MDEV-23855}{MDEV-23855} made InnoDB scan an ever
longer list of closed tablespaces whenever the open-file limit
(\texttt{innodb\_open\_files}) was reached; with tens of thousands of tables,
SHOW TABLE STATUS became hundreds of times slower, and a dump of 600 schemas grew
from 2:45 to 27 minutes until fix \mariacommit{3f75d8b2} kept open tablespaces at the
head of the list. In DuckDB, commit \duckcommit{19795040} rewrote the Top-N operator
around a heap of sort keys, and queries with a large OFFSET or LIMIT took about
6.9\,s instead of 0.86\,s until fix \duckcommit{8e2c944a} brought them to 0.77\,s.
In CockroachDB, whose catalog descriptors are read from the key-value layer,
release 23.1 also fetched comments and zone configurations by point lookups, so
a pg\_catalog introspection query from Django's test suite took 1--2\,s instead
of milliseconds; fix \crdbcommit{eb8bdebe} switched to range scans. None of these
regressions changes a query plan, and each depends on a dimension that an
ordinary benchmark holds fixed: the number of tables, the OFFSET, or the size of
the catalog.

Accepted trade-offs are also unevenly spread. They make up 44.4\% of
session/protocol families (8/18), 35.0\% of schema/catalog families (14/40), and
29.6\% of types/expressions families (8/27), but only about one tenth of query
execution (13/124) and storage I/O (12/122) families. For example, TiDB pull
request \idref{https://github.com/pingcap/tidb/pull/26905}{\#26905} made a USE statement apply to the following statements of a
multi-statement request, and its authors declared that the fix consumes more CPU
because it disables the point-plan cache in some edge cases. A detector for the
former modules will often report slowdowns that maintainers accepted, so it must
separate accepted costs from defects (Section~\ref{sec:accepted-costs}).

\finding{RQ1}{Regressions occur throughout the kernel. Query optimization is the
largest module (27.4\%), but query execution and storage I/O together account
for more (36.8\%). The leading module depends on the architecture: query
optimization in embedded and distributed systems, storage I/O in client/server
systems, and query execution in analytical engines. Testing plan changes alone
cannot cover these regressions.}

\subsection{RQ2: Causal Mechanisms}
\label{sec:rq2}

\begin{figure}[t]
\centering
\includegraphics[width=\columnwidth]{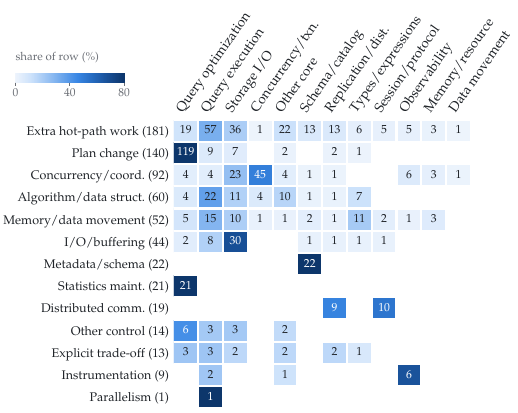}
\caption{Primary mechanism (rows) by primary module (columns) of the 668 kernel
families. Numbers are family counts; shading is the share within each row.
Blank cells are zero.}
\label{fig:mechanisms}
\end{figure}

Figure~\ref{fig:mechanisms} reports the normalized change mechanisms. The largest
class is extra or repeated work on a hot path (181, 27.1\%). It includes new
checks inside tuple loops, redundant expression evaluation, and repeated
metadata or system calls. Plan-selection or transformation changes follow (140,
21.0\%), then concurrency or coordination changes (92, 13.8\%). These
three mechanisms cover 413 families (61.8\%).

A module label does not determine the mechanism. Plan changes are concentrated,
with 119 of 140 (85.0\%) in query optimization, whereas extra hot-path work
appears in all 12 modules, led by query execution (57) and storage I/O (36).
Concurrency or coordination changes split between concurrency and transactions
(45) and storage I/O (23). In the other direction, 64 of the 183
optimizer-module families (35.0\%) are not plan changes; statistics maintenance
(21) and hot-path work (19) are the largest of these. Mechanisms also vary with
architecture. Concurrency or coordination changes make up 21.0\% of
client/server families but only one SQLite family, and all 22
metadata/schema-maintenance and all 19 distributed-communication families come
from TiDB. A statistics change can add optimizer-time work and alter the
eventual plan, and a synchronization change can affect latency, throughput, and
CPU at once, so a module label alone is a weak guide to test strategy. A
detector needs predictions for the module, the mechanism, the likely metric, and
the triggering dimensions.

\begin{figure}[t]
\centering
\setlength{\fboxsep}{0pt}\setlength{\fboxrule}{0.4pt}%
\begin{lrbox}{\diffbox}%
\fontsize{6.8}{8}\selectfont
\setlength{\tabcolsep}{3pt}\arrayrulecolor{black!35}%
\begin{tabular}{p{\dimexpr0.70\columnwidth-6.4pt\relax}>{\raggedleft\arraybackslash}p{\dimexpr0.30\columnwidth-6.4pt\relax}}
\diffhead{\textbf{Introducing commit} \pgcommit{9ca67658d}}{execExpr.c: ExecBuildHash32Expr()} \\
\diffdel\lstinline[style=codeline]| scratch.resvalue = &state->resvalue;| &  \\
\diffdel\lstinline[style=codeline]| scratch.resnull = &state->resnull;| &  \\
\diffadd\lstinline[style=codeline]| if (i == list_length(hash_exprs) - 1) {| &  \\
\diffadd\lstinline[style=codeline]|   scratch.resvalue = &state->resvalue;| & \codenote{final key} \\
\diffadd\lstinline[style=codeline]|   scratch.resnull = &state->resnull;| &  \\
\diffadd\lstinline[style=codeline]| } else {| &  \\
\diffadd\lstinline[style=codeline]|   scratch.resvalue = &iresult->value;| & \codenote{earlier keys} \\
\diffadd\lstinline[style=codeline]|   scratch.resnull = &iresult->isnull;| &  \\
\diffadd\lstinline[style=codeline]| }| &  \\
\diffhead{\textbf{Fix} \pgcommit{60826a352d}}{execExprInterp.c: EEOP\_HASHDATUM\_FIRST\_STRICT} \\
\diffctx\lstinline[style=codeline]| if (fcinfo->args[0].isnull) {| &  \\
\diffdel\lstinline[style=codeline]|   *op->resnull = true;| & \codenote{earlier key: iresult only} \\
\diffdel\lstinline[style=codeline]|   *op->resvalue = (Datum) 0;| &  \\
\diffadd\lstinline[style=codeline]|   state->resnull = true;| & \codenote{expression is NULL} \\
\diffadd\lstinline[style=codeline]|   state->resvalue = (Datum) 0;| &  \\
\diffctx\lstinline[style=codeline]|   EEO_JUMP(op->d.hashdatum.jumpdone);| &  \\
\diffctx\lstinline[style=codeline]| }| &  \\
\hline
\multicolumn{2}{>{\raggedright\arraybackslash}p{\dimexpr\columnwidth-6.8pt\relax}}{\sffamily\textbf{Impact:} with at least ten
equality keys and frequent NULLs, 3.5\,ms before, 20,843\,ms after (5,948$\times$),
and 3.4\,ms after the fix.} \\
\end{tabular}%
\end{lrbox}%
\fcolorbox{black!35}{white}{\usebox{\diffbox}}
\caption{A PostgreSQL hash-join regression and its fix. A NULL in an earlier
join key marks only the intermediate result NULL, so the tuple enters the
hash table but never matches. The fix also covers \texttt{NEXT32\_STRICT}.
Gray notes are ours; we omit comments and an assertion.}
\label{fig:hashjoin}
\end{figure}

Four cases illustrate the range:

\noindent\textbf{Plan transformation.} SQLite check-in \sqlitecommit{198b3e33} changed LEFT
JOIN view flattening. For the preserved reproducer, the regression replaced a
primary-key lookup with a full scan and increased virtual-machine steps by
about 69.5$\times$; fixing check-in \sqlitecommit{dc9f025d} restored the plan.

\noindent\textbf{Input-sensitive hot path.} PostgreSQL change \pgcommit{9ca67658d} caused a
multi-key hash semi join to retain ineffective tuples when a non-final key was
NULL (Figure~\ref{fig:hashjoin}). With high NULL frequency and at least ten equality keys, local measurements
show 3.504\,ms before the change, 20,843.143\,ms after it, and 3.439\,ms after
fix \pgcommit{60826a352d}---a 5,948$\times$ regression relative to baseline. The
trigger is far narrower than a generic join benchmark.

\noindent\textbf{Algorithmic storage behavior.} PostgreSQL commit \pgcommit{b2a5545}
introduced repeated \texttt{stat()} calls while recycling WAL segments, yielding
$O(N^2)$ behavior in the number of segments. Fix \pgcommit{5ae1572993} changed the
maintenance path. Here, neither SQL equivalence nor a plan oracle is sufficient;
the directed test must scale WAL-segment count and observe system calls or
maintenance latency.

\noindent\textbf{Code layout.} Layout itself can be causal: PostgreSQL change
\pgcommit{6150a1b0} enlarged \texttt{BufferDesc} on some platforms so instances
crossed a 64-byte cache-line boundary, producing false sharing. The fixing change
\pgcommit{008608b9d5} removed an extra spinlock. This regression requires a
platform- and concurrency-sensitive microbenchmark; query-plan comparison gives
no useful signal.

\finding{RQ2}{Three mechanisms cause most regressions: extra work on a hot
path, plan selection or transformation, and concurrency or coordination, which
together account for 61.8\% of families. Plan changes mostly stay within the
optimizer, whereas hot-path work appears in every module, so no single change
signal or workload covers them.}

\subsection{RQ3: Evidence Completeness}

Table~\ref{tab:evidence} separates evidence types. Public evidence references are
available for 600 families (89.8\%); 511 (76.5\%) identify a fixing revision.
Code localization is stronger: 644 (96.4\%) map to changed files and 640 (95.8\%)
to enclosing functions or diff hunks. However, only 292 families (43.7\%) have an
explicit introducing revision, and only 155 (23.2\%) have both introducing and
fixing references. At least one of the two revisions is available for 648
families (97.0\%).

\begin{table}[t]
\caption{Evidence completeness by system. Values are counts, not percentages.}
\label{tab:evidence}
\centering
\scriptsize
\setlength{\tabcolsep}{3.1pt}
\begin{tabular}{lrrrrrr}
\toprule
System & Kernel & Intro. & Fix & Public & Files & Func./hunk \\
\midrule
SQLite      & 87  & 82 & 37  & 62  & 78  & 77  \\
PostgreSQL  & 143 & 69 & 132 & 100 & 143 & 140 \\
DuckDB      & 35  & 23 & 30  & 35  & 35  & 35  \\
TiDB        & 136 & 85 & 72  & 136 & 136 & 136 \\
MySQL       & 102 & 12 & 76  & 102 & 87  & 87  \\
MariaDB     & 41  & 4  & 40  & 41  & 41  & 41  \\
CockroachDB & 32  & 12 & 32  & 32  & 32  & 32  \\
OceanBase   & 27  & 0  & 27  & 27  & 27  & 27  \\
ClickHouse  & 65  & 5  & 65  & 65  & 65  & 65  \\
\midrule
Total       & 668 & 292& 511 & 600 & 644 & 640 \\
\bottomrule
\end{tabular}
\end{table}

Trigger conditions are explicitly structured for 503 families (75.3\%). All 668
have an evidence-derived targeted-test hypothesis. Given this asymmetry, the
full corpus is suitable for module/mechanism analysis and many
forms of fix localization, while only the 292-family subset provides sound
ground truth for introducing-commit prediction.

Introducing revisions are also unevenly available. They are known for 60.7\% of
plan changes (85/140) but only 23.2\% of extra hot-path work (42/181), for
83.3\% of accepted trade-offs (100/120) but 34.2\% of upstream-confirmed
regressions (185/541), and for 94.3\% of SQLite families but 29.7\% of
client/server families. The 292-family subset thus differs from the corpus. Plan changes rise from 21.0\% to 29.1\% and accepted
trade-offs from 18.0\% to 34.2\%, whereas extra hot-path work falls from 27.1\%
to 14.4\%. An evaluation limited to known introducing commits overweights optimizer regressions and SQLite, and should report results per
mechanism or reweight them to the corpus. The dynamic evaluation in
Section~\ref{sec:dynamic} inherits this skew.

\finding{RQ3}{Most families can be localized in code, but fewer than half can
be traced to the change that caused them. Over 95\% map to changed files or
functions, whereas only 43.7\% identify the introducing revision, and this
subset over-represents plan changes and accepted trade-offs. Commit-level evaluations need ground truth chosen by task, results reported per
mechanism, and fixes never treated as causes.}

\newcommand{\DynamicSelected}{60}
\newcommand{\DynamicValidated}{13}
\newcommand{\DynamicInconclusive}{18}
\newcommand{\DynamicUnpackaged}{29}

\section{Dynamic Evaluation}
\label{sec:dynamic}

This section answers RQ4. Section~\ref{sec:dynamic-design} builds executable
evaluation cases from the corpus, Section~\ref{sec:dynamic-tools} describes the
detectors and how their alerts are attributed to a regression family, and
Section~\ref{sec:dynamic-results} reports what they find and miss.

\subsection{Evaluation Cases}
\label{sec:dynamic-design}

To evaluate existing detectors, we derive executable evaluation cases from the
162 confirmed families whose introducing diff resolves in our source mirrors,
excluding explicit performance trade-offs. Extending a 20-family pilot, we first
select a predeclared target of 60 candidate families in 60 distinct
introducing-change groups across eight systems and ten mechanisms, before any
check of build or runtime feasibility. Predeclared mechanism quotas keep large
classes from dominating: ten each for hot-path work, plan changes, and
concurrency; eight each for algorithms and memory; six for I/O; four for
statistics; two for other control changes; and one each for metadata and
parallelism, the only eligible family in each. Within these quotas, we spread
candidates across databases and trigger conditions.

Each case pins the introducing commit, its first parent, and the linked fix when
one exists. A primary workload exercises the reported trigger, a negative
control does not, and both are fixed with their expected results before
measurement. After warmup, at least five interleaved ABBA blocks compare the
parent and introducing revisions, with a fixing-revision run after each block,
and we bootstrap complete blocks to obtain a 95\% interval for the relative
slowdown. A case passes reference acceptance when its results are correct, its
timings exceed the timer floor, its primary interval lies entirely at or above
5\%, its control interval lies entirely below 5\%, and the fixing revision is
faster again. Observations without a linked fix therefore remain inconclusive.

Of the \DynamicSelected{} candidates, \DynamicValidated{} pass acceptance,
\DynamicInconclusive{} remain inconclusive, and \DynamicUnpackaged{} lack
executable fixtures. Candidates fail mainly because the control slows as
well, because no usable fix shows recovery, or because our host cannot recreate
the historical comparison; for example, one PostgreSQL regression depends on
spinlock size on particular architectures. Untimed system-call counts and
EXPLAIN plans check that reference cases reach their target code paths.

Because this selection precedes any feasibility check, we then check all 162 families for whether their historical versions build and run, and measure 24 more. In total, 55 families have measurements and 31 pass reference acceptance, including 2 whose workloads are scaled-down adaptations of the original reports. From these, we choose 22 evaluation cases in distinct change groups, covering all 9 in-scope mechanisms with unequal quotas: six planning, four algorithm, four hot-path, three memory, and one in each of the other five categories. The quotas cap planning cases because 10 of the 12 accepted planning references come from SQLite. By system, the cases comprise 12 SQLite, 5 DuckDB, 4 PostgreSQL, and 1 TiDB. Statistics maintenance is excluded as a standalone category, and distributed communication and instrumentation lack eligible cases.

\subsection{Detectors and Fix-Boundary Attribution}
\label{sec:dynamic-tools}

\newcommand{\VfourCalibrated}{15}
\newcommand{\VfourCases}{19}
\newcommand{\VthreeAlerts}{787}
\newcommand{\VthreeAttributed}{220}
\newcommand{\VfourNegatives}{62}
\newcommand{\VfourNegativeAlerts}{0}
\newcommand{\VfourPgSqlsmithExcluded}{86.7}
\newcommand{\VfourDuckBaseRejected}{93.6}

\begin{table}[t]
\centering\footnotesize
\caption{Generator outcomes of SQLancer~2.0.0, one configuration per system, and SQLsmith on the 22 cases; Figure~\ref{fig:tool-outcomes} covers all configurations. Fix-boundary attribution is retrospective for SQLite
and predeclared for DuckDB and PostgreSQL (PG). ``Signature, boundary'' counts calibrated cases
with a reproduced reference signature and with a valid fix boundary.}
\label{tab:generator-outcomes}
\setlength{\tabcolsep}{3pt}
\begin{tabular}{lrrrr}
\toprule
Outcome & SQLite & DuckDB & PG & TiDB \\
\midrule
Selected cases & 12 & 5 & 4 & 1 \\
Signature, boundary (of) & 9, 10 (10) & 3, 3 (5) & 3, 4 (4) & -- \\
\midrule
Family-attributed alert & 6 & 1 & 2 & 0 \\
Unattributed alerts only & 6 & 0 & 1 & 0 \\
Alerts, invalid boundary & 0 & 2 & 0 & 0 \\
No alert, calibrated & 8 & 2 & 2 & 0 \\
No alert, uncalibrated & 0 & 0 & 1 & 0 \\
Tool incompatible & 0 & 0 & 2 & 0 \\
Not applicable & 4 & 5 & 0 & 2 \\
\bottomrule
\end{tabular}
\end{table}

We run every detector that we could run on the historical versions. The SQL
generators are SQLancer, with its NoREC, TLP~\cite{rigger2020tlp},
PQS~\cite{rigger2020pqs}, and CODDTest~\cite{zhang2025coddtest} oracles and its
DuckDB and PostgreSQL generators, and SQLsmith. APOLLO~\cite{jung2019apollo} and
AMOEBA~\cite{liu2022amoeba} target performance, so each runs its own
detection loop for 900 seconds per trial; APOLLO runs with three declared fixes
to its released driver, and AMOEBA's rewriter works only on its own demo schema.
Benchmark replay runs the developers' benchmarks, SQLite's speedtest1, DuckDB's
micro-benchmarks, and pgbench, on every version of each case. We could not run
QPG~\cite{ba2023qpg}, which mutates the database that our protocol keeps fixed,
or the unreleased PUPPY~\cite{wu2025puppy}, Hulk~\cite{wu2025hulk}, and
QueryZen~\cite{ba2026queryzen}. Counting each SQLancer oracle and generator
separately gives nine detector configurations, and we call each pairing of a
case with a configuration or benchmark a \emph{combination}.

Every statement that a tool generates or reports is replayed on the parent and
introducing revisions. The evaluator keeps statements whose typed results agree
across versions and confirms each timing alert in 30 interleaved blocks;
same-binary negative controls raise \VfourNegativeAlerts{} alerts in
\VfourNegatives{} pairs, so measurement noise alone rarely produces an alert.
Combinations that a tool cannot target (not applicable) or on which it fails
are reported apart from misses: two SQLite
triggers need write transactions that read-only replay excludes, SQLancer cannot
set up the TiDB case, SQLsmith has no DuckDB or TiDB backend, and
SQLancer~2.0.0 cannot initialize the two oldest PostgreSQL snapshots.

\noindent\textbf{Fix-boundary attribution.} A timing alert shows that the introducing
revision is slower than its parent, not that the slowdown belongs to the studied
family; many SQLite alerts are broad slowdowns across query shapes unrelated to
the reported trigger. We therefore re-measure each alert in 30 blocks across the
fix and its first parent and attribute it to the family only if this boundary
removes at least half of the introducing slowdown on a log scale, with a
bootstrap lower bound of at least 5\%. Each case's reference workload must pass
the same test; otherwise its boundary is invalid. For SQLite, we defined these
criteria after the first runs and applied them retrospectively, attributing
\VthreeAttributed{} of \VthreeAlerts{} alerts; for DuckDB and PostgreSQL, they
were predeclared.

\subsection{Results}
\label{sec:dynamic-results}

\begin{figure}[t]
\centering
\includegraphics[width=\columnwidth]{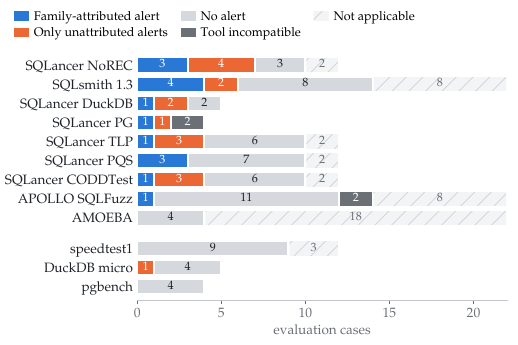}
\caption{Outcomes per detector and benchmark-replay baseline over the 22 cases.
Each bar counts the case combinations in the tool's target systems by outcome:
a family-attributed alert, only unattributed alerts, no alert, a tool failing
on the historical version, or not applicable. TLP, PQS, and CODDTest use the
current SQLancer main branch; the other SQLancer rows use release 2.0.0. No tool
applies to the TiDB case.}
\label{fig:tool-outcomes}
\end{figure}

Table~\ref{tab:generator-outcomes} and Figure~\ref{fig:tool-outcomes} summarize
the outcomes. Calibration reproduces \VfourCalibrated{} of \VfourCases{} reference
signatures. Across the nine detector configurations, \VfiveGeneratorFamily{} of
\VfiveGeneratorApplicable{} applicable combinations, covering
\VfiveGeneratorFamilyCases{} cases, contain family-attributed alerts, and the
developers' benchmarks attribute none of \VfiveBaselineApplicable{}. The clearest
hit is the SQLite LEFT JOIN view case, where SQLancer and SQLsmith generate LEFT
JOINs over the view that run 119--527 times slower, with a plan change that the
fix reverts. Using a lexical proxy of each reference trigger construct, we also
check whether a tool's statements ever contain the construct. The tools fail for
different reasons:
\begin{itemize}
\item \textbf{SQLancer} (six bars in Figure~\ref{fig:tool-outcomes}, 49
applicable combinations). Ten
combinations, covering six cases, contain family-attributed alerts. PQS's pivot-row
checks also wrap queries in a FROM-clause
subquery and slow the view co-routine case by 84--411\%. Other attributions are
broad: TLP and CODDTest succeed only in the covering-index case, whose change
slows table scans whether or not queries touch its virtual columns. In 27 of
its 37 misses, no statement contains the trigger construct: on SQLite, SQLancer
never emits JSON functions, \texttt{matchinfo}, recursive CTEs, or ORDER BY with
LIMIT, and on PostgreSQL it never calls the fixture's SQL function. Release 2.0.0 also cannot initialize the two 9.x PostgreSQL snapshots,
and the base binder rejects \VfourDuckBaseRejected\% of its DuckDB statements.
\item \textbf{SQLsmith} (14 applicable combinations on SQLite and PostgreSQL).
Four combinations contain family-attributed alerts, including a PostgreSQL SQL
function that the reference also calls, which slows by 27\%. In 8 of its 10
misses, no statement contains the trigger construct; for example, its
PostgreSQL DML never targets the parent of the 1,001-table inheritance tree. In
the other two, the construct appears in the wrong form: in the hash-join case
of Figure~\ref{fig:hashjoin}, its 101 EXISTS semi-joins lack the many
NULL-bearing equality keys. On PostgreSQL, \VfourPgSqlsmithExcluded\% of its statements are excluded
because they use unseeded TABLESAMPLE or functions that depend on randomness,
the clock, or the session.
\item \textbf{APOLLO} (14 applicable combinations). Its own detection rules
discard most candidates before our evaluator sees them. On SQLite,
\VfiveApolloSqliteAbove{} of \VfiveApolloSqliteTimed{} timed queries exceed its
twofold threshold, but \VfiveApolloSqliteBelowMinimum{} of these run below its
10\,ms minimum on our fixtures; its only adjudicated SQLite alert, in the
join-order case, is attributed but differs in shape from the reference. On
PostgreSQL, its 1.1-fold rule on single EXPLAIN ANALYZE timings yields
\VfiveApolloPgReports{} reports on the two 9.x snapshots, none of which survives
repeated measurement, and its query filter skips the UPDATE statements that the
inheritance case needs. Its 2018 SQLsmith fork fails on the two newer
snapshots.
\item \textbf{AMOEBA} (four PostgreSQL cases). It forms \VfiveAmoebaSets{} sets
of equivalent queries and reports \VfiveAmoebaReports{} pairs whose members
differ in time, but none yields a confirmed cross-version alert. Its oracle
compares formulations within one version rather than across versions, and its
rewriter works only on its own demo schema, which contains none of the fixture
objects that the triggers involve.
\item \textbf{Developer benchmarks} (\VfiveBaselineApplicable{} applicable
combinations). None attributes a regression; the largest slowdowns are
\VfiveSpeedtestMax\% in speedtest1 and \VfivePgbenchMax\% in pgbench. Their
workloads are fixed: pgbench runs single-table select-only and TPC-B-like
transactions on its own schema, and speedtest1 batches its writes, so it misses
even the in-memory case, whose regression adds time to every commit. DuckDB's
micro-benchmarks raise one alert, a \VfiveMicroAlertEffect\% slowdown that the
fix does not remove, although its budget ran only
\VfiveMicroShareLow--\VfiveMicroShareHigh\% of them in the four larger cases.
\end{itemize}

\finding{RQ4}{Mostly not. The detectors raise many timing alerts, but
re-measuring them across the fixing commit attributes alerts to the studied
regression in only \VfiveGeneratorFamilyCases{} of the 22 cases, and the
developers' benchmarks attribute none. The detectors miss mainly because their workloads never contain the triggering
construct; timer resolution explains few misses.}

\section{Discussion}
\label{sec:discussion}

\subsection{Comparison with Prior Findings}

Our results agree with general performance-bug studies on where time is lost
but differ in what database kernels add. Jin et al.\ attribute 76 of 109
performance bugs in five applications to uncoordinated or skippable function
calls~\cite{jin2012performance}; extra or repeated work on a hot path is likewise
our largest mechanism (27.1\%). Plan selection or transformation, our second
mechanism (21.0\%), has no counterpart in their categories or in the six root
causes that Chen and Shang identify for Hadoop and
RxJava~\cite{chen2017exploratory}, because a database can slow down when a
change alters which plan it chooses, without any change to the code that executes the
plan. Jin et al.\ also report that synchronization causes 22\% of server bugs
but 4.4\% of client bugs. We see the same split between architectures, with concurrency or coordination
changes causing 21.0\% of client/server families and only one SQLite family. Their finding that almost half of the bugs need inputs
with both special features and large scales also holds for our corpus: 47.8\% of
families need a larger data scale, and each of them also needs at least one
other test dimension, such as SQL shape or schema (Table~\ref{tab:dimensions}).

Chen and Shang find that most regression-introducing commits in Hadoop and
RxJava fix other bugs, and Ren et al.\ trace most slowdowns of Linux core
operations to a few security enhancements, new features, and
misconfigurations~\cite{ren2019linux}. Our 120 accepted trade-offs (18.0\%) show the same pressure in database kernels, where maintainers knowingly accept a slowdown
for correctness, security, or another goal. Unlike these measurement-driven studies, our corpus contains only regressions that developers confirmed
or accepted, so it describes regressions that mattered enough to be
acknowledged, not every measurable slowdown.

\subsection{Implications for Database Developers}

Two results concern how projects test and record regressions. First, the
developers' own benchmarks attribute none of the 22 evaluated regressions
(Section~\ref{sec:dynamic-results}). Fixed suites such as speedtest1 and pgbench
exercise common paths, whereas the regressions in our corpus need a particular
query shape, data distribution, table count, or catalog size
(Section~\ref{sec:results}). When a change touches a known risk area, a
microbenchmark that varies the corresponding dimension of
Table~\ref{tab:mapping} likely costs less than diagnosing the regression after
release. Second, the evidence that later analysis needs is recorded unevenly. An
introducing revision is known for 82 of 87 SQLite families and 85 of 136 TiDB
families, but for 4 of 41 MariaDB families, 5 of 65 ClickHouse families, and none
of the 27 OceanBase families (Table~\ref{tab:evidence}). Recording the
introducing commit when a regression is fixed would make it usable for
bisection, attribution, and detector evaluation. Accepted costs are likewise
visible mainly where a process asks for them: TiDB's pull-request template has a
side-effect field for CPU and memory regressions, and 62 of TiDB's 64 accepted
trade-offs, more than half of the 120 in the corpus, were declared there.

\subsection{Implications for Detector Design}
\label{sec:implications}

\begin{table*}[t]
\caption{Evidence-derived mapping from change signals to directed tests and oracles.}
\label{tab:mapping}
\centering
\small
\renewcommand{\arraystretch}{1.2}
\hyphenpenalty=10000
\newcolumntype{L}[1]{>{\raggedright\arraybackslash}p{#1}}
\newcolumntype{Y}{>{\raggedright\arraybackslash}X}
\begin{tabularx}{\textwidth}{L{3.4cm}L{4.2cm}YY}
\toprule
Diff or module signal & Principal risk & Targeted workload dimensions & Preferred oracle \\
\midrule
Cost model, cardinality, rewrite rule & Plan loss or unfavorable plan & SQL shape, schema, statistics, skew, selectivity & Plan/operator structure, estimate error, latency \\
\rowcolor{black!4}Join, aggregate, sort, expression operator & Extra work or algorithmic degradation & Cardinality, row width, types, NULL rate, selectivity & Operator time, CPU/instructions, throughput \\
Buffer, pager, WAL, checkpoint, file format & I/O amplification or cache-policy degradation & Working set, cache state, file layout, write ratio & Bytes/IOPS, cache misses, tail latency \\
\rowcolor{black!4}Lock, latch, atomic, task scheduler & Contention, serialization, scheduling overhead & Concurrency, conflict rate, transaction length & Throughput, lock wait, CPU, P95/P99 \\
Allocation, copying, memory layout & Allocation/copy amplification or poor locality & Row width, batch size, cardinality, memory limit & Allocations, peak memory, cache misses \\
\rowcolor{black!4}Catalog, DDL, statistics maintenance & Metadata scan or maintenance amplification & Table/column/partition count, concurrent DDL & DDL latency, scanned rows, lock wait \\
RPC, shard/region routing, retry policy & Request amplification or network wait & Topology, shard count, batch size, fault injection & RPC/retry count, network bytes, latency \\
\rowcolor{black!4}Logging, tracing, metrics & Instrumentation entering a hot path & Feature switches, QPS, statement complexity & CPU, allocations, throughput, event count \\
\bottomrule
\end{tabularx}
\end{table*}

Table~\ref{tab:mapping} turns recurring evidence patterns into requirements for
a change-directed detector, which (1) analyzes a diff at file, function, and
call-path levels; (2) predicts module and mechanism risks; (3) selects trigger
dimensions; (4) instantiates targeted tests; and (5) combines structural signals
with robust measurements. Our results constrain each step.

\noindent\textbf{Predict mechanisms, not only modules.} Because plan changes mostly stay in the optimizer while hot-path work spans every module (RQ2), a file path says little about how a change can slow the kernel. A change to structure layout, loop
placement, early exits, locking, or a cost formula can share vocabulary with
harmless maintenance. A detector should consider control-flow and
call-frequency changes, data-structure and synchronization operations, and
optimizer-rule semantics, and rank changed hunks with plausible mechanisms
instead of one module label.

\begin{table}[t]
\caption{Most frequent derived test dimensions (non-exclusive).}
\label{tab:dimensions}
\centering
\scriptsize
\setlength{\tabcolsep}{3.3pt}
\begin{tabular}{lrr@{\hspace{12pt}}lrr}
\toprule
Dimension & $n$ & \% & Dimension & $n$ & \% \\
\midrule
Data scale        & 319 & 47.8 & Operator select.    & 124 & 18.6 \\
SQL shape         & 307 & 46.0 & Storage state        & 122 & 18.3 \\
Schema            & 223 & 33.4 & I/O layout           & 122 & 18.3 \\
Statistics        & 183 & 27.4 & Concurrency          & 69  & 10.3 \\
Data distribution & 183 & 27.4 & Configuration        & 62  & 9.3  \\
Contention        & 51  & 7.6  & Transaction pattern  & 51  & 7.6  \\
\bottomrule
\end{tabular}
\end{table}

\noindent\textbf{Vary the suspected trigger.} Table~\ref{tab:dimensions} lists the
non-exclusive dimensions that the corpus's test hypotheses require. Data scale
and SQL shape are the most common, but neither dominates. Schema, statistics,
and data distribution mediate plan choices, storage state and I/O layout matter
for 18.3\% of families, and concurrency-related dimensions form a smaller but
operationally distinct region. The detectors in Section~\ref{sec:dynamic} miss
mainly because their workloads never contain the trigger. A change near a
hash-join NULL path should therefore drive key count and NULL placement, a
cache-line layout change thread count and pinning, and a WAL loop the segment
count, rather than leave the trigger to a fixed workload.

\noindent\textbf{Combine structural and timing oracles.} Timing alone reports slowdowns outside the studied family and can miss the
change itself. Most timing alerts in Section~\ref{sec:dynamic} are version
differences outside the studied family. Conversely, in the WAL-recycling case,
\texttt{stat} calls grow from 255 in the base to 8,510 in the candidate and drop
to 252 at the fix, yet checkpoint latency does not recover at that fix. A
detector should pair latency, throughput, CPU, memory, I/O, lock-wait, and RPC
measurements with structural evidence such as plans, operator counts, and system
calls, and should base a verdict on repeated interleaved runs and effect sizes
rather than on one elapsed-time comparison.

\subsection{Implications for Evaluating Detectors}
\label{sec:accepted-costs}

Three practices follow from RQ3 and the dynamic evaluation. First, ground truth
must match the task and respect time. The 292 families with explicit introducing
changes can be ordered by commit time: older families may support rule or model
construction, and newer families form a holdout. Test templates, taxonomy
refinements, and code representations derived from a held-out fix must not be
exposed to the detector, and because this subset over-represents plan changes
and accepted trade-offs, results should be reported per mechanism or reweighted
to the corpus. Second, an alert should count only if it belongs to the studied
regression. Re-measuring it across the fixing commit
(Section~\ref{sec:dynamic-tools}) decides this independently of the mechanism,
but requires both an introducing revision and a linked fix. Third, accepted
trade-offs need their own label. A correctness or security fix may intentionally
make a narrow path slower; a detector should still quantify the cost and the
affected workload, but a policy layer, not the performance oracle, decides
whether it blocks a change. Training on trade-offs as ordinary bugs would teach a detector to flag intended costs, and discarding them would remove 18.0\% of the corpus. Suitable metrics include Recall@$K$ and mean
reciprocal rank for ranking risky commits or hunks, trigger rate and generation
success for tests, false-positive rate and stable effect size for the oracle, and
time or resource savings relative to full benchmark replay.

\section{Threats to Validity}
\label{sec:threats}

\noindent\textbf{Internal validity.} Our confirmation rule relies on explicit upstream
causality, and an upstream statement can be incomplete; we retain the evidence
so that readers can audit each causal chain. We reproduced only seven families
locally, because reproducing every family would require historical toolchains,
hardware, datasets, and deployments and would favor newer, simpler cases, while
an acknowledged fix or revert already establishes the cause. The
corpus therefore supports analyses of code causes and triggers, not comparisons
of slowdown magnitude. Deterministic normalization and a single primary category
can misclassify cross-cutting changes. In the
dynamic evaluation, all measurements come from one shared host and have not
been replicated elsewhere, and the
fixed 5\% threshold may miss smaller effects. Fix-boundary attribution assumes
that the linked fix is the complete repair, and it shows that an alert belongs
to the family, not that it reproduces the reported symptom. The SQLite
attribution criteria and trigger proxies were defined after its first runs,
whereas those for DuckDB and PostgreSQL were declared before measurement.
APOLLO and AMOEBA run through our adapters, which differ from their original
setups.

\noindent\textbf{External validity.} The selected systems cover four broad architecture
styles, but not closed-source services or every open-source engine, and
differences in language, governance, and deployment affect transferability.
Projects expose different channels and historical depths, and the corpus misses
regressions that were never reported, were discussed privately, or left no
performance terms in their public record; multi-source search and cross-link
expansion reduce but cannot eliminate this bias. Because candidate and source
counts differ in units, they cannot yield project-level defect rates, so we
report per-system composition rather than universal incidence. The corpus is a
snapshot ending August 30, 2026, and later reports and fixes can change a
family's evidence. The 22 evaluation cases were chosen by mechanism quotas and
executability rather than at random, so they favor embedded engines, with 17 of
the 22 on SQLite or DuckDB, and cover distributed and concurrency regressions
only sparsely. Seeded tool runs disclose schema and data derived from known
references and use small smoke-test budgets, so they do not measure
unconstrained discovery or support a tool ranking.

\section{Conclusion}
\label{sec:conclusion}

We studied 668 publicly confirmed database-kernel performance regression
families from nine open-source systems. Query optimization is the largest module, but execution, storage, and
concurrency together account for more; hot-path work, plan changes, and
coordination are the leading causal mechanisms; and source localization is much
more available than the true introducing revision, which is known
far more often for plan changes. These findings support a
change-directed approach that classifies diff semantics, selects conditional
workload dimensions, and combines structural and multi-metric oracles, and the
corpus provides task-specific ground truth for developing and evaluating it,
provided accepted trade-offs, fixing changes, and introducing changes remain
distinct. Evaluating nine detector configurations, including APOLLO and AMOEBA,
and three developer benchmarks on 22 evaluation cases derived from the corpus
yields two lessons for evaluating change-directed detectors. First, a timing alert across the introducing commit does not by itself
rediscover the regression, because most alerts are broad version differences
outside the studied family; re-measuring each alert across the fixing commit
attributes it to the family independently of the mechanism. Second, most misses are generation gaps: in the corresponding cases the
generators never emit the triggering construct, replaying the developers'
benchmarks attributes none of the regressions, timer sensitivity explains few
misses, and several cases need writes or session state that read-only replay
excludes.

\ifCLASSOPTIONcaptionsoff
  \newpage
\fi

\bibliographystyle{IEEEtran}
\bibliography{references}

\end{document}